\documentclass[a4paper,conference]{IEEEtran}
\usepackage{cite}
\ifCLASSINFOpdf
\else
\fi
\ifCLASSOPTIONcompsoc
 \usepackage[caption=false,font=normalsize,labelfont=sf,textfont=sf]{subfig}
\else
 \usepackage[caption=false,font=footnotesize]{subfig}
\fi
\usepackage{url}

\usepackage[english]{babel}

\usepackage{xcolor}
\usepackage{soul}
\usepackage[bookmarks=false,hidelinks]{hyperref}

\usepackage{cite}
\usepackage{amsmath,amssymb,amsfonts}
\usepackage{algorithmic}
\usepackage{graphicx}
\usepackage{textcomp}
\usepackage[capitalize]{cleveref}
\usepackage{flushend}

\usepackage[range-phrase={\,--\,}, range-units=single, per-mode=symbol, detect-all]{siunitx}

\makeatletter
  \@addtoreset{paragraph}{section}
\makeatother

\usepackage{booktabs}
\usepackage{glossaries-extra}
\setabbreviationstyle[common]{short-nolong}
\setabbreviationstyle[long]{long-noshort}
\newabbreviation[category={common}]{3d}{3D}{dreidimensional}
\newabbreviation[category={common}]{3gpp}{3GPP}{3rd Generation Partnership Project}
\newabbreviation[category={common}]{ai}{AI}{Artificial Intelligence}
\newabbreviation{api}{API}{application programming interface}
\newabbreviation{ar}{AR}{Augmented Reality}
\newabbreviation[category={common}]{cad}{CAD}{computer-aided design}
\newabbreviation{cam}{CAM}{Class Activation Map}
\newabbreviation{drl}{DRL}{deep reinforcement learning}
\newabbreviation{erp}{ERP}{Enterprise Resource Planning}
\newabbreviation{faas}{FaaS}{Function-as-a-Service}
\newabbreviation{fec}{FEC}{forward error correction}
\newabbreviation{ga}{GA}{general aviation}
\newabbreviation[category={common}]{gps}{GPS}{Global Positioning System}
\newabbreviation[category={common}]{gpu}{GPU}{graphics processing unit}
\newabbreviation{h3d}{H3D}{holographic 3D}
\newabbreviation{hmd}{HMD}{head-mounted display}
\newabbreviation{iot}{IoT}{Internet of Things}
\newabbreviation{isv}{ISV}{independent software vendor}
\newabbreviation[category={common}]{it}{IT}{Information Technology}
\newabbreviation{jcas}{JCAS}{Joint Communications and Sensing}
\newabbreviation{kpi}{KPI}{Key Performance Indicator}
\newabbreviation{lidar}{LIDAR}{Light Detection and Ranging}
\newabbreviation{lstm}{LSTM}{long short term memory}
\newabbreviation{mcs}{MCS}{Modulation and Coding Scheme}
\newabbreviation{mr}{MR}{Mixed Reality}
\newabbreviation{ml}{ML}{Machine Learning}
\newabbreviation{moq}{MoQ}{Media over QUIC}
\newabbreviation{nbmp}{NBMP}{network-based media processing}
\newabbreviation[category={common}]{oran}{O-\glsxtrshort{ran}}{Open \glsxtrshort{ran}}
\newabbreviation{poc}{PoC}{proof of concept}
\newabbreviation{pubsub}{Pub/Sub}{Publish/Subscribe}
\newabbreviation{qoe}{QoE}{Quality of Experience}
\newabbreviation{qos}{QoS}{Quality of Service}
\newabbreviation{ran}{RAN}{radio access network}
\newabbreviation[firstplural=radio access technologies]{rat}{RAT}{radio access technology}
\newabbreviation{redcap}{RedCap}{Reduced Capabilities}
\newabbreviation{rem}{REM}{radio environment map}
\newabbreviation[category={common}]{sdk}{SDK}{Software Development Kit}
\newabbreviation{sdr}{SDR}{software-defined radio}
\newabbreviation{slam}{SLAM}{Simultaneous Localization and Mapping}
\newabbreviation{tcas}{TCAS}{traffic alert and collision avoidance system}
\newabbreviation{uav}{UAV}{unmanned aerial vehicle}
\newabbreviation{vr}{VR}{Virtual Reality}
\newabbreviation{xr}{XR}{eXtended Reality}
\newabbreviation{6g}{6G}{sixth-generation}
\newabbreviation{csi}{CSI}{channel state information}
\newabbreviation{usrp}{USRP}{Universal Software Radio Peripheral}
\newabbreviation{oai}{OAI}{Openairinterface5g}
\newabbreviation{nic}{NIC}{network interface card}
\newabbreviation{los}{LoS}{line-of-sight}
\newabbreviation{rssi}{RSSI}{received signal strength indicator}
\newabbreviation{cfr}{CFR}{channel frequency response}
\newabbreviation{srs}{SRS}{sounding reference signal}
\newabbreviation{pusch}{PUSCH}{physical uplink shared channel}
\newabbreviation{5g}{5G}{fifth-generation}
\newabbreviation{gru}{GRU}{gated recurrent unit}
\newabbreviation{rnn}{RNN}{recurrent neural network}
\newabbreviation{snr}{SNR}{signal to noise ratio}
\newabbreviation{bler}{BLER}{block error rate}
\newabbreviation{gNB}{gNB}{next Generation NodeB}
\newabbreviation{ue}{UE}{user equipment}
\newabbreviation{cqi}{CQI}{channel quality index}
\newabbreviation{harq}{HARQ}{hybrid automatic repeat request}
\newabbreviation{cpe}{CPE}{common phase error}

\usepackage{balance}
\usepackage[all]{nowidow}

\usepackage{graphics}
\graphicspath{{figures/}}

\title{Empowering AI towards 6G: Realistic UAV Channel Data Acquisition using Open-Source Solutions}
\author{Intelligente Netze }
\date{November 2025}

\usepackage{eso-pic}
\AddToShipoutPictureBG*{%
  \AtPageUpperLeft{%
    \raisebox{-1.15cm}{\makebox[\paperwidth][c]{\parbox{0.86\paperwidth}{\centering\scriptsize
    This is the author's accepted version of a paper presented at the 2026 IEEE 103rd Vehicular Technology Conference (VTC2026-Spring).\\
    \copyright~2026 IEEE. Personal use of this material is permitted. Permission from IEEE must be obtained for all other uses, in any current or future media, including reprinting/republishing this material for advertising or promotional purposes, creating new collective works, for resale or redistribution to servers or lists, or reuse of any copyrighted component of this work in other works.}}}%
  }%
}

\begin{document}

\author{
    \IEEEauthorblockN{
        Qiuheng Zhou\IEEEauthorrefmark{1}, Sergiy Melnyk\IEEEauthorrefmark{1}, Hans D. Schotten\IEEEauthorrefmark{1}\IEEEauthorrefmark{2},
        Robert Vilter\IEEEauthorrefmark{3},
        Nick Stuckert\IEEEauthorrefmark{3}
    }
    \IEEEauthorblockA{
        \IEEEauthorrefmark{1}
        Intelligent Networks,
        German Research Center for Artificial Intelligence (DFKI GmbH), Kaiserslautern, Germany \\
        Email: \{qiuheng.zhou, sergiy.melnyk, hans\_dieter.schotten\}@dfki.de
    }
    \IEEEauthorblockA{
        \IEEEauthorrefmark{2}
        Institute for Wireless Communication and Navigation,
        University of Kaiserslautern-Landau, Kaiserslautern, Germany\\
        Email: schotten@rptu.de
    }
    \IEEEauthorblockA{
        \IEEEauthorrefmark{3}
        Aviation Engineering Group, Technical University of Applied Sciences Wildau, Wildau, Germany \\
        Email: \{vilter, stuckert\}@th-wildau.de
    }
}

\maketitle

\begin{abstract}
The ``sim-to-real'' gap introduced by hardware impairments motivates realistic experimental platforms for AI-native research in Non-Terrestrial Networks (NTN). This paper presents an open-source UAV-to-Ground measurement testbed built on the OpenAirInterface (OAI) stack and commodity Software Defined Radios (SDRs), enabling direct extraction of raw Sub-6 GHz Channel State Information (CSI) synchronized with flight telemetry. Through real-world flight campaigns, we observe a velocity-bandwidth decoupling effect: under unguided, measurement-only ridge extraction, dominant spectral ridges deviate strongly from the theoretical kinematic Doppler trajectory, whereas guided tracking substantially reduces this mismatch. This indicates that Doppler information is present, but not dominant, in the raw observable dynamics of commodity SDR-based measurements. We further benchmark representative deep learning predictors on this hardware-inclusive dataset and show, through a case study, that their behavior differs substantially from idealized expectations. By open-sourcing the testbed and dataset, we provide a realistic foundation for training and evaluating AI models for practical aerial network deployments.
\end{abstract}


\glsresetall
\section{Introduction}\label{section1}
\Gls{6g} wireless networks are envisioned to support novel applications demanding unprecedented performance in terms of data rates, latency, and reliability~\cite{10494372,9403380,10054381}. Achieving these ambitious goals requires a paradigm shift towards intelligent network management, where \gls{ai} plays a central role. A critical component for training and validating these \gls{ai} models is the availability of accurate and realistic \gls{csi}.

Recent studies have leveraged \gls{ai} methods to process historical \gls{csi} for predicting future channel states, achieving high accuracy in simulated fast-fading environments~\cite{10422880,9832933,9044427, 10757983}. Nevertheless, rigorous performance assessment requires strict evaluation and real-world validation before deployment. While over-the-air simulation software has evolved significantly, it inherently simplifies the complex stochastic nature of physical hardware. Conversely, evaluations using commercial \gls{5g} equipment are often restricted by ``black-box'' limitations, preventing access to low-level channel estimates. Although dedicated channel sounders have been used to characterize vehicular and \gls{uav} channels~\cite{8653275,10616106}, they typically provide sterile, offline measurements. However, these methods typically fail to capture the real-time performance impacts of authentic system characteristics. A major oversight is hardware-induced phase noise: in practical open-source SDR systems, common phase fluctuations can dominate the observable spectral ridge and obscure the direct manifestation of purely kinematic Doppler effects assumed in idealized simulations. As a result, models trained only on idealized data may generalize poorly to real receivers.

This creates an urgent need for flexible, open-source experimentation platforms capable of \textit{in-situ} measurement. Unlike external packet sniffing or expensive proprietary sounders, an ideal framework should capture the \gls{gNB}'s internal channel estimates directly during active data transmission, preserving the authentic artifacts of the hardware RF chain. To address the scarcity of such datasets, particularly for dynamic \gls{uav} scenarios, this paper introduces a comprehensive framework for acquiring real-world \gls{5g} channel data.

The main contributions of this paper are threefold. First, we introduce an open-source \gls{uav}-\gls{5g} measurement framework based on \gls{oai}~\cite{8727207} and \glspl{usrp}. This system bridges simulation and practice by extracting raw complex-valued channel estimates synchronized with telemetry. Second, we release a multi-modal dataset that provides empirical evidence that, in practical SDR-based links, hardware-induced phase fluctuations can dominate the observable spectral ridge and obscure a direct kinematic Doppler interpretation.\footnote{UAV 5G CSI dataset: \url{https://git-in.dfki.de/Greatqz/uav-csi}} Third, we benchmark representative prediction models on the collected hardware-inclusive CSI and show that model behavior differs substantially from idealized expectations, highlighting the need to train and evaluate channel predictors on realistic measured data rather than simulation-only channels.

\section{System Architecture and Implementation}\label{section2}
In this section, we detail the proposed open-source measurement framework. Unlike commercial ``black-box'' systems, our platform is architected as a fully programmable standalone \gls{5g} network, enabling direct access to physical layer parameters essential for \gls{ai} training. The system comprises two primary subsystems: a software-defined base station and a custom-built \gls{uav} \gls{ue}.

\begin{figure}[t!]
    \centering
    \subfloat[System-level Architecture]{
        \includegraphics[width=0.74\linewidth]{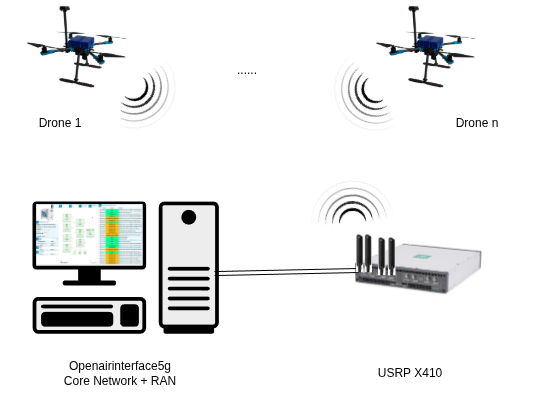}
        \label{fig:arch_diagram}
    }
    
    \vspace{0.05cm}
    
    \subfloat[UAV Prototype Hardware]{
        \includegraphics[width=0.74\linewidth]{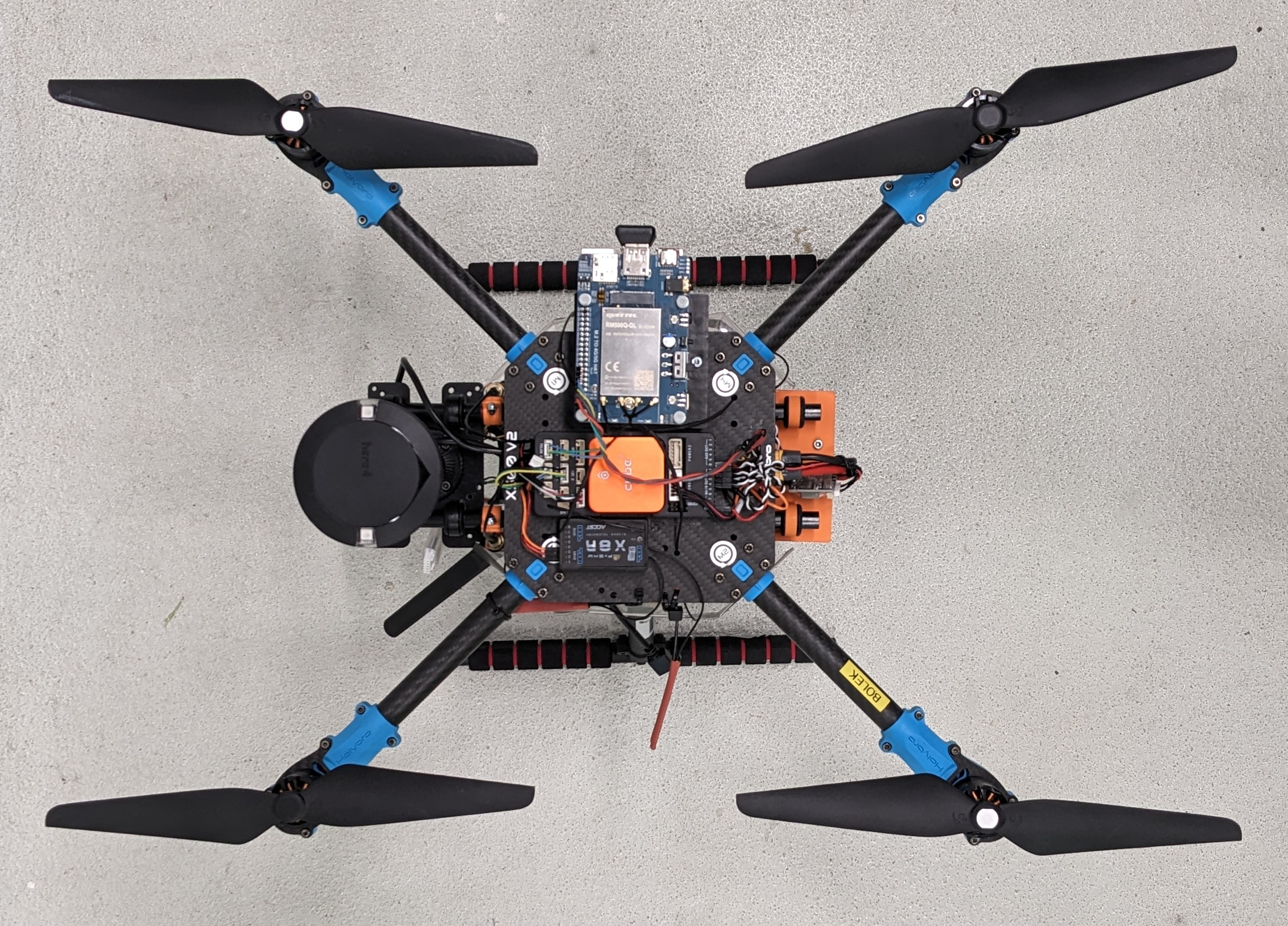}
        \label{fig:uav_photo}
    }
    \caption{The proposed open-source measurement framework. (a) The system spans from the OAI-based 5GC/RAN to the UAV UE. (b) The custom UAV prototype integrating the 5G modem and flight controller.}
    \label{fig:sys_overview}
\end{figure}

The high-level architecture and the UAV prototype are illustrated in Fig.~\ref{fig:sys_overview}.

\subsection{Base Station and CSI Extraction}
On the network side (Fig.~\ref{fig:arch_diagram}), a high-performance PC hosts both the \gls{5g} Core (5GC) and the \gls{5g} \gls{ran}. The basic configurations adhere to the official \gls{oai} tutorials\footnote{NR SA Tutorial OAI CN5G: \url{https://gitlab.eurecom.fr/oai/openairinterface5g/-/blob/develop/doc/NR_SA_Tutorial_OAI_CN5G.md}}\footnote{NR SA Tutorial COTS UE: \url{https://gitlab.eurecom.fr/oai/openairinterface5g/-/blob/develop/doc/NR_SA_Tutorial_COTS_UE.md}}. Based on this setup, the \gls{sdr} \gls{ran} software performs all digital baseband processing and interfaces with a \gls{usrp} X410 via a \SI{10}{GbE} \gls{nic}. The \gls{usrp} functions as the RF front-end, upconverting the digital I/Q samples to broadcast the \gls{5g} NR signal at a carrier frequency of \SI{3619.2}{\mega\hertz}. This setup provides stable coverage up to \SI{300}{\meter} under \gls{los} conditions.

\begin{figure}[t!]
    \centering
    \includegraphics[width=\linewidth]{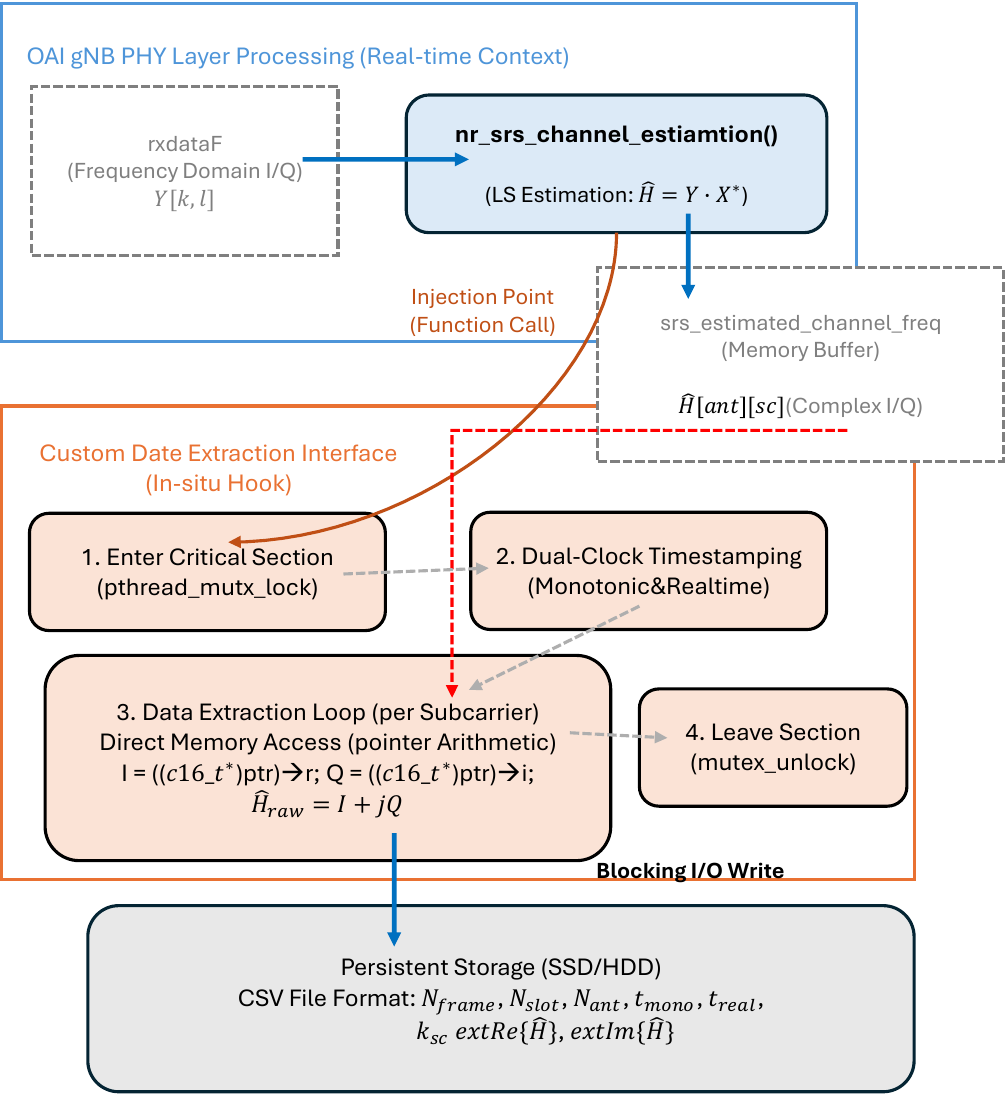}
    \caption{SRS-based CSI acquisition workflow. The instrumented \gls{oai} gNB estimates per-subcarrier complex CSI from the uplink \gls{srs} and logs the synchronized output for post-flight analysis.}
    \label{fig:csi_pipeline}
\end{figure}

To bridge the ``sim-to-real'' gap, we instrumented the \gls{oai} uplink \gls{srs} processing chain. During sounding, the \gls{ue} transmits a known \gls{srs} reference sequence, and the \gls{oai} uplink estimator correlates the received signal with this sequence to obtain the complex-valued channel response for each subcarrier and antenna port. Our modification taps this estimator output immediately after \gls{srs}-based channel estimation, before higher-layer reporting or compression, thereby exposing the same internal \gls{csi} representation that practical prediction algorithms would observe at the receiver. Fig.~\ref{fig:csi_pipeline} summarizes this workflow. A thread-safe mutex-based logger preserves data integrity without disrupting live operation.

Using pre-programmed SIM cards, the system supports up to 16 \glspl{uav} simultaneously connecting to the base station, with an aggregate uplink throughput of up to \SI{130}{\mega\bit\per\second}.

\subsection{UAV Prototype and Data Synchronization}
The aerial platform is a Holybro X500 multicopter (Fig.~\ref{fig:uav_photo}), integrating an Orange Cube flight controller (running ArduPilot) and a Raspberry Pi 4B. The Raspberry Pi bridges the 5G modem and flight controller, ensuring synchronization between CSI and telemetry.

A critical challenge in multi-modal data acquisition is alignment. While the flight controller logs high-frequency inertial telemetry, the \gls{gps} updates occur at a lower rate of \SI{1}{\hertz}. In our framework, this \gls{gps} telemetry serves a role beyond positioning: it provides the ground truth for temporal segmentation. By synchronizing the high-resolution \gls{csi} timestamps with GNSS UTC time, we can slice the continuous data stream into distinct flight phases.

\section{Data Measurement and Analysis}\label{section3}
To support \gls{ai}-driven research with high-fidelity data, our framework captures the full complex-valued \gls{cfr}. The measurement campaign was conducted at Airport Schönhagen, Germany, accumulating a total flight time of approximately \SI{100}{\minute} across varied mission profiles. The resulting dataset comprises more than \num{600000} temporal snapshots of wideband channel estimates (Table~\ref{tab:dataset_summary}), supporting experimental benchmarking and prototype evaluation.

\begin{table}[htbp]
\caption{Summary of the Collected \gls{uav}-\gls{5g} Channel Dataset}
\begin{center}
\begin{tabular}{c cccc} 
\toprule 
\textbf{Mission} & \textbf{Altitude} & \textbf{Max. Velocity} & \textbf{Sampling Rate} & \textbf{\gls{gps} Rate} \\
\midrule 
1 & \SI{40}{\meter} & \SI{5}{\meter\per\second} & \SI{10}{\milli\second} (\SI{100}{\hertz}) & \SI{1}{\hertz} \\
2 & \SI{40}{\meter} & \SI{10}{\meter\per\second} & \SI{10}{\milli\second} (\SI{100}{\hertz}) & \SI{1}{\hertz} \\
3 & \SI{40}{\meter} & \SI{15}{\meter\per\second} & \SI{10}{\milli\second} (\SI{100}{\hertz}) & \SI{1}{\hertz} \\
\hline
4 & \SI{50}{\meter} & \SI{5}{\meter\per\second} & \SI{5}{\milli\second} (\SI{200}{\hertz}) & \SI{1}{\hertz} \\
5 & \SI{50}{\meter} & \SI{10}{\meter\per\second} & \SI{5}{\milli\second} (\SI{200}{\hertz}) & \SI{1}{\hertz} \\
\hline
6 & \SI{60}{\meter} & \SI{10}{\meter\per\second} & \SI{10}{\milli\second} (\SI{100}{\hertz}) & \SI{1}{\hertz} \\
7 & \SI{60}{\meter} & \SI{15}{\meter\per\second} & \SI{10}{\milli\second} (\SI{100}{\hertz}) & \SI{1}{\hertz} \\
\bottomrule 
\end{tabular}
\label{tab:dataset_summary}
\end{center}
\end{table}

\subsection{Spatial-Domain Characterization}
\begin{figure}[t!]
    \centering
    \includegraphics[width=0.8\linewidth]{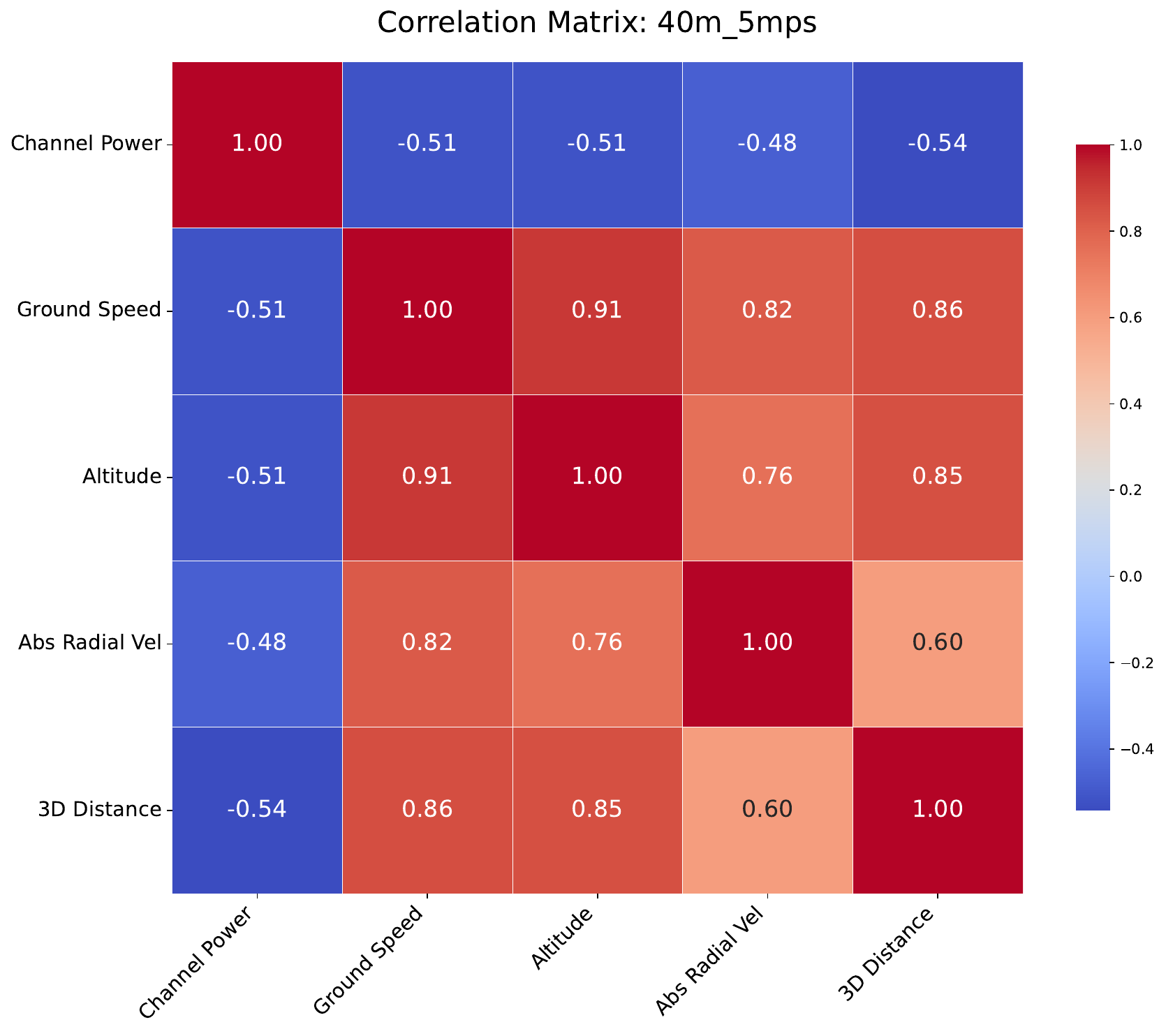}
    \caption{Pearson correlation matrix of key kinematic and channel variables.}
    \label{fig:corr_matrix}
\end{figure}

We first examine the channel behavior in the spatial domain to identify large-scale fading characteristics associated with the UAV's geometric position. Representative trajectories show clear fading zones as the \gls{uav} moves toward boundary regions far from the gNB. To quantify these dependencies, we calculated the Pearson correlation matrix (Fig.~\ref{fig:corr_matrix}). The results show a strong negative correlation ($-0.54$) between channel power and the 3D distance to the gNB, which is consistent with path-loss-related attenuation. Furthermore, channel power negatively correlates with groundspeed ($-0.51$) and absolute radial velocity ($-0.48$), suggesting that stronger platform dynamics are associated with lower received power in this measurement set.

\subsection{Time-Domain Dynamics}
\begin{figure}[t!]
    \centering
    \includegraphics[width=\linewidth]{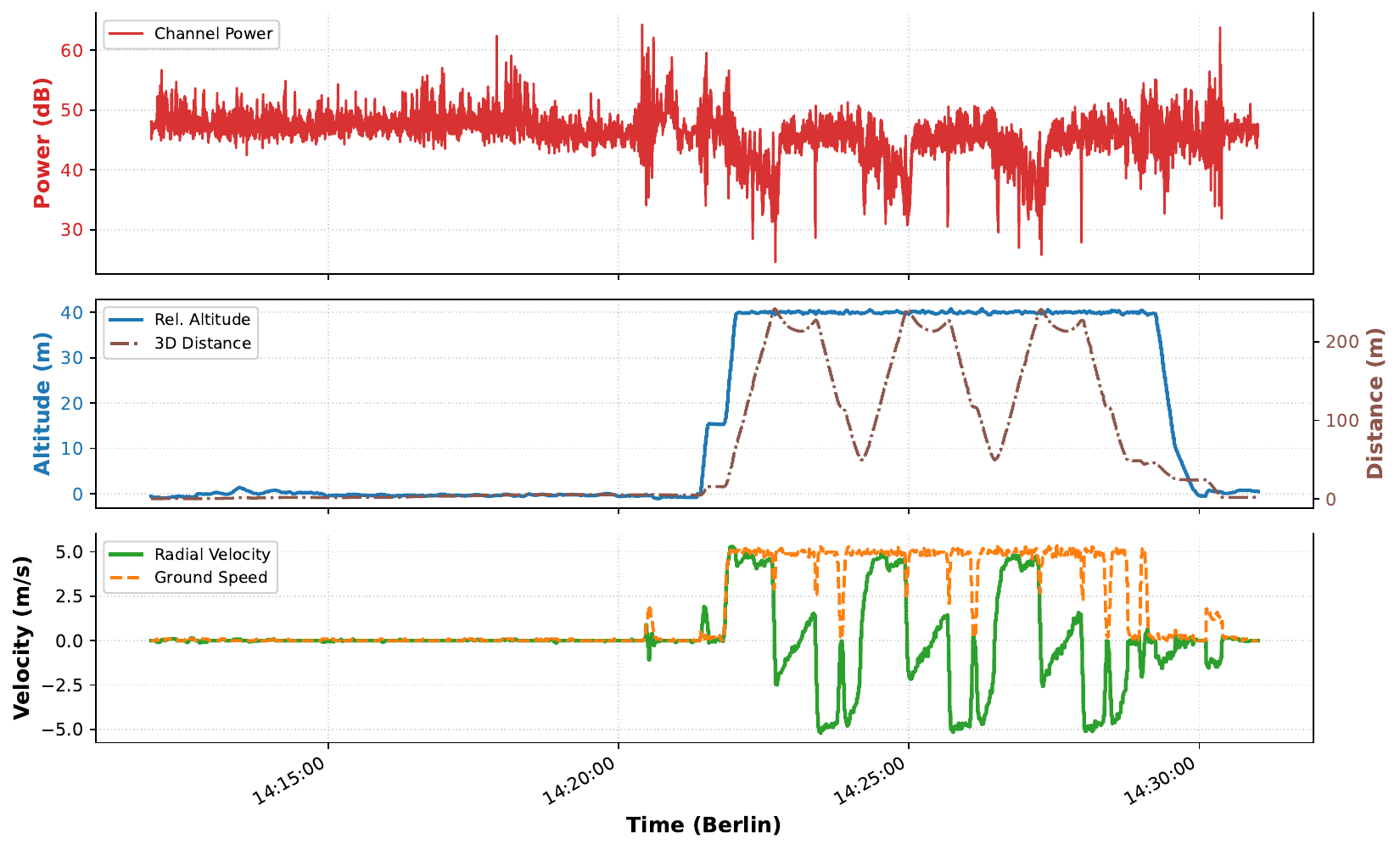}
    \caption{Time-series profile synchronizing channel power with telemetry.}
    \label{fig:time_profile}
\end{figure}

Having established the spatial correlations, we next investigate the temporal evolution of the channel to understand its stability under motion. Fig.~\ref{fig:time_profile} synchronizes the instantaneous channel power with flight telemetry. A clear temporal association can be observed: the average power trend fluctuates in response to altitude and heading changes. Notably, rapid deep fades are superimposed on this trend, highlighting the fast-varying nature of the channel and the resulting challenge for feedback loops.

\begin{figure}[t!]
    \centering
    \includegraphics[width=\linewidth]{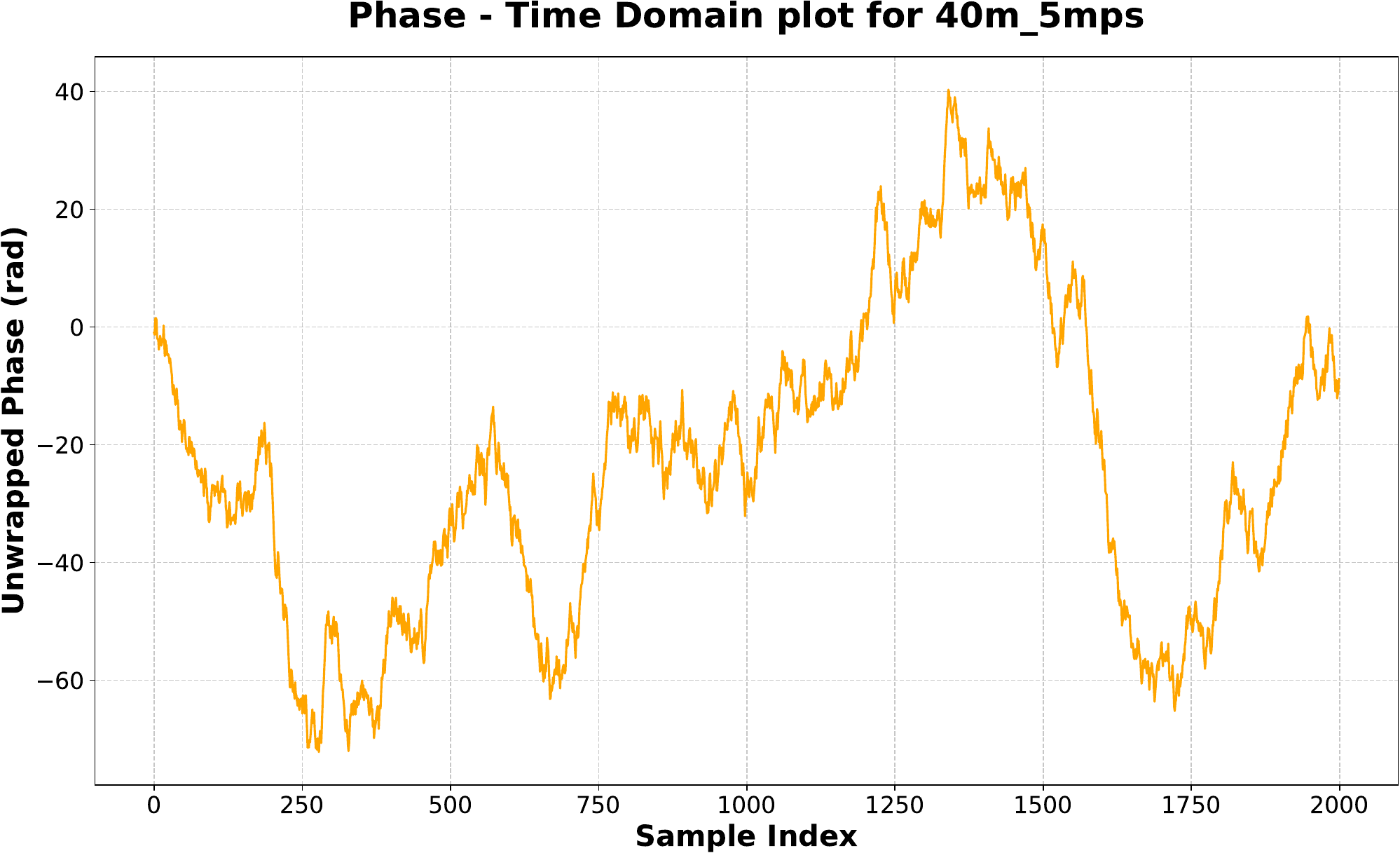}
    \caption{Unwrapped phase response distinguishing motion-induced Doppler drift from hardware phase noise.}
    \label{fig:time_phase}
\end{figure}

Crucially, the phase response (Fig.~\ref{fig:time_phase}) provides insight into signal coherence. The trajectory exhibits a superposition of a slow drift component consistent with Doppler/CFO-like behavior and rapid jitter. This jitter is a signature of hardware phase noise inherent in commodity \glspl{sdr}. Capturing this impairment is vital, as it is often absent in synthetic datasets but critically affects the performance of \gls{ai}-based channel predictors.

\subsection{Frequency-Domain Structure}
\begin{figure}[t!]
    \centering
    \includegraphics[width=\linewidth]{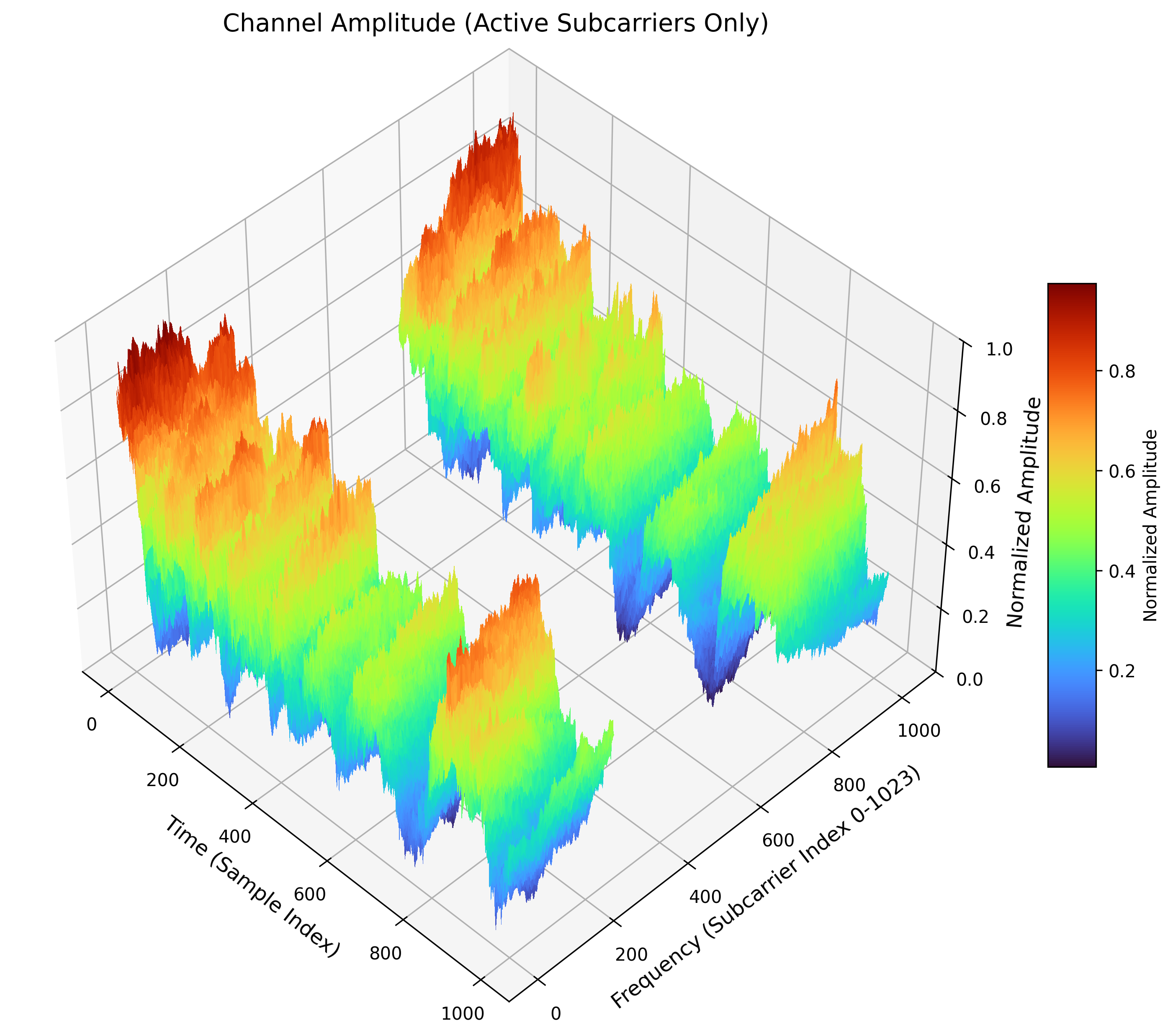}
    \caption{3D surface plot of raw channel estimates, illustrating dual time-frequency selectivity.}
    \label{fig:srs_surface}
\end{figure}

The wideband channel estimates also exhibit pronounced frequency selectivity in addition to their temporal variation. In other words, the measured \gls{csi} evolves as a genuinely two-dimensional time-frequency process rather than as a flat-fading sequence, which motivates predictors capable of learning dependencies across both dimensions. Fig.~\ref{fig:srs_surface} shows localized fading valleys and ridges jointly across time and subcarrier index, underscoring that realistic channel prediction in this setting is intrinsically a 2D inference problem.

\subsection{Spectral Analysis}
\begin{figure}[t!]
    \centering
    \includegraphics[width=\linewidth]{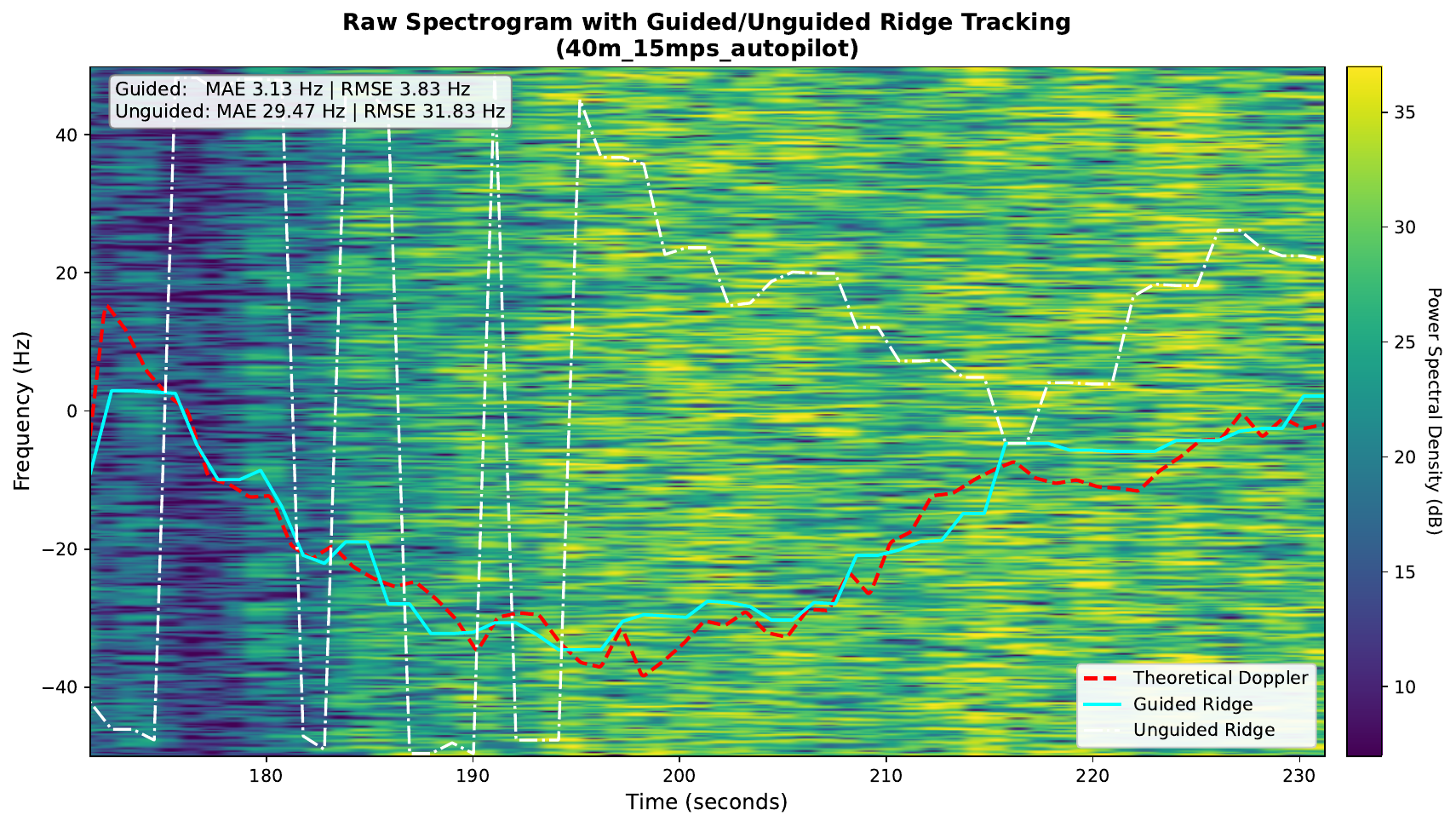}
    \caption{Illustrative phase-2 excerpt of the high-speed mission (\SI{15}{m/s}) with theoretical Doppler (red dashed), guided ridge (cyan), and unguided ridge (white).}
    \label{fig:spectrogram_15mps}
\end{figure}

To investigate the impact of flight dynamics on the channel, we analyze an illustrative phase-2 excerpt from the high-speed flight (Mission No.~3, \SI{15}{m/s}) in Fig.~\ref{fig:spectrogram_15mps}, where the raw spectrogram is overlaid with the theoretical Doppler trajectory and the extracted guided and unguided ridges. The displayed \SI{60}{s} window is selected using a fixed measurement-side criterion that favors large guided-versus-unguided separation together with substantial theoretical Doppler sweep. The theoretical maximum Doppler shift at this velocity is approximately \SI{181.0}{Hz}. Given the system sampling rate of \SI{100}{Hz}, motion-induced components at this speed should undergo frequency folding within the observable baseband.

The key observation is that the unguided, measurement-only ridge does not consistently follow the expected aliased Doppler trajectory. Instead, the dominant spectral energy remains concentrated near the center of the measured bandwidth, and the extracted unguided ridge is attracted to this hardware-dominated structure. Over the displayed excerpt, the unguided track yields \SI{29.47}{Hz} MAE and \SI{31.83}{Hz} RMSE with respect to the theoretical Doppler curve, whereas guided tracking reduces the error to \SI{3.13}{Hz} MAE and \SI{3.83}{Hz} RMSE.

This comparison shows that Doppler information is present in the measurement, but it is not the dominant component of the raw observable ridge. Relying solely on raw spectral maxima can therefore misrepresent the underlying kinematic behavior.

\section{Benchmark Evaluation}\label{section4}

In this section, we utilize standard deep learning architectures as a benchmark to quantify the impact of the hardware impairments identified in Section~\ref{section3} on channel state prediction. The objective is to evaluate whether common models, trained on real-world \gls{sdr}-based \gls{csi}, can distinguish between physical fading and hardware-induced artifacts. This evaluation serves to bridge the ``sim-to-real'' gap by highlighting the necessity of hardware-inclusive datasets for training robust AI models in practical deployments.

\subsection{Rationale and Problem Formulation}
Conventional channel prediction heavily relies on \glspl{rnn} to capture sequential dependencies. However, we hypothesize that the recurrent inductive bias of these architectures, which inherently assumes smooth, Markovian state transitions, becomes a bottleneck in the presence of the significant hardware jitter identified in Section~\ref{section3}. 

Unlike \glspl{rnn}, the Transformer’s self-attention enables global dependency modeling, functioning as an adaptive filter against hardware-induced jitter \cite{10757983}. In this context, the attention map effectively functions as a non-linear, adaptive low-pass filter. By globally aggregating information, the model can identify the consistent phase drift associated with kinematic Doppler shifts while suppressing the incoherent, high-frequency spectral broadening caused by hardware phase noise. This selective attention mechanism allows the Transformer to learn a robust representation of the fading envelope, mitigating the ``sim-to-real'' mismatch where models otherwise overfit to hardware-specific artifacts.

The goal is to predict the future complex channel vector $\mathbf{h}_{t+\tau}$ given a sequence of historical estimates. Let $\mathbf{H}_{in} = [\mathbf{h}_{t-L+1}, \dots, \mathbf{h}_t]$ denote the input sequence of length $L$, where each $\mathbf{h}_k \in \mathbb{C}^{N_{sc}}$ represents the frequency-domain channel estimate at time $k$ across $N_{sc}$ subcarriers. The model $f(\cdot)$ aims to minimize the prediction error $\hat{\mathbf{h}}_{t+\tau} = f(\mathbf{H}_{in}; \theta)$, where $\tau$ is the prediction horizon and $\theta$ represents the trainable parameters.

\subsection{Experimental Setup}
\subsubsection{Dataset Selection}
We use the \SI{15}{\meter\per\second} cruising subset from Section~\ref{section3}, split chronologically into training (70\%), validation (15\%), and testing (15\%) sets. The observation window is $T_{\text{obs}} = 10$ channel snapshots, and the evaluated horizons are $k = 1$ to $10$; Fig.~\ref{fig:prediction_result} shows the $k = 5$ case.

\subsubsection{Baselines and Architecture}
We compare the Transformer against two baselines: Naive (Last-Value), which assumes $\hat{\mathbf{h}}_{t+k} = \mathbf{h}_t$, and gated recurrent unit (\gls{gru}) as the representative \gls{rnn} baseline. For a fair architectural comparison, both trainable models use two layers with hidden dimension 512; the Transformer further employs 16 attention heads.

\subsubsection{Implementation Details}
The predictor input is the magnitude of the complex \gls{csi} across 576 active subcarriers, standardized on the training split only. Training uses Adam ($10^{-5}$, batch size 512) with validation-based early stopping, typically converging within 150 epochs. All results follow this chronological train/validation/test protocol.

\subsection{Prediction Performance: Handling Phase Noise}
\subsubsection{Qualitative Analysis}
\begin{figure}[t!]
    \centering
    \includegraphics[width=\linewidth]{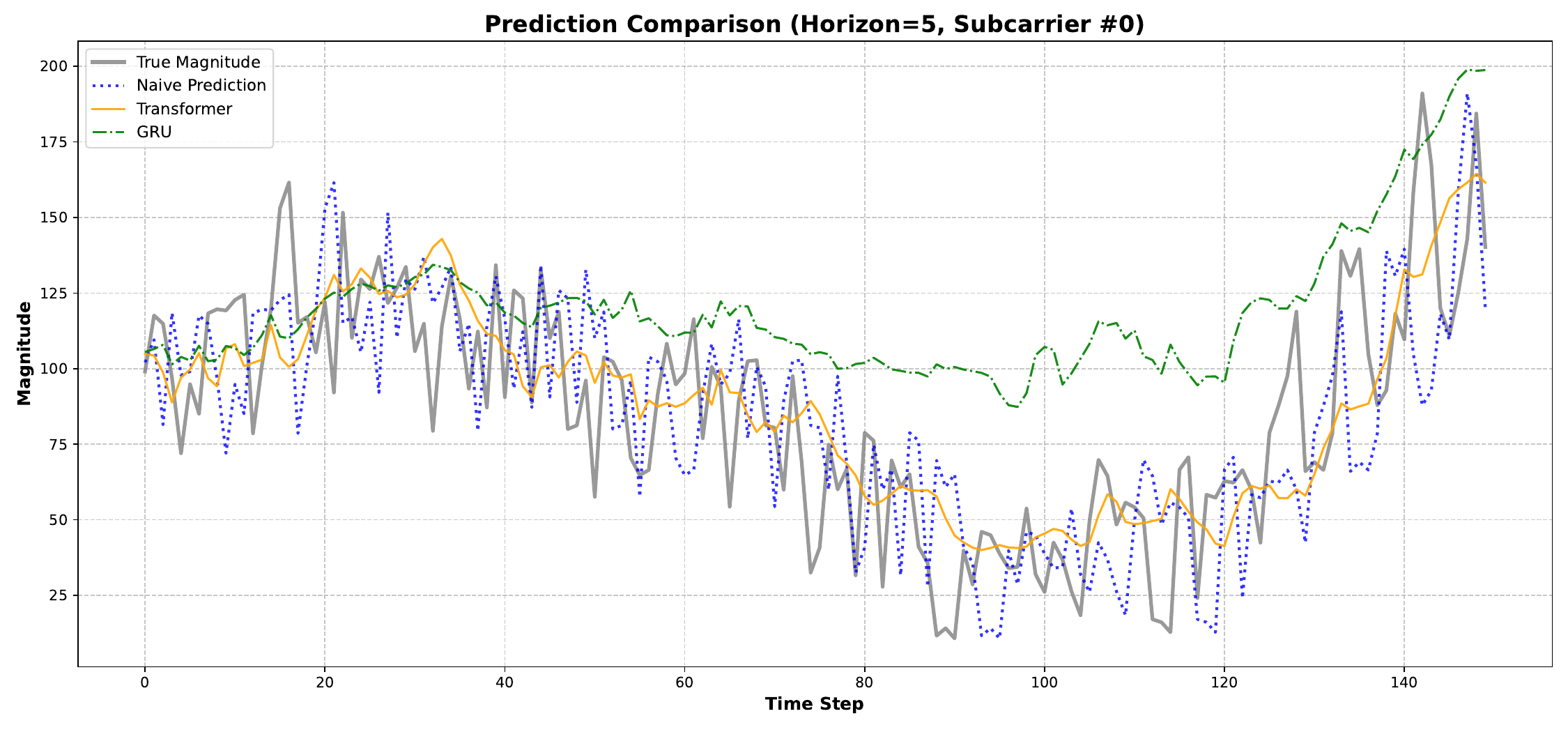}
    \caption{Qualitative comparison (Horizon=5). The Transformer tracks the fading trend while filtering jitter, whereas the GRU over-smooths.}
    \label{fig:prediction_result}
\end{figure}

Fig.~\ref{fig:prediction_result} illustrates the $k = 5$ prediction case. The \gls{gru} (green) over-smooths the trajectory and misses local extrema, whereas the Transformer (orange) better preserves the fading trend while filtering high-frequency hardware jitter.

\subsubsection{Quantitative Analysis}
\begin{figure}[t!]
    \centering
    \includegraphics[width=0.92\linewidth]{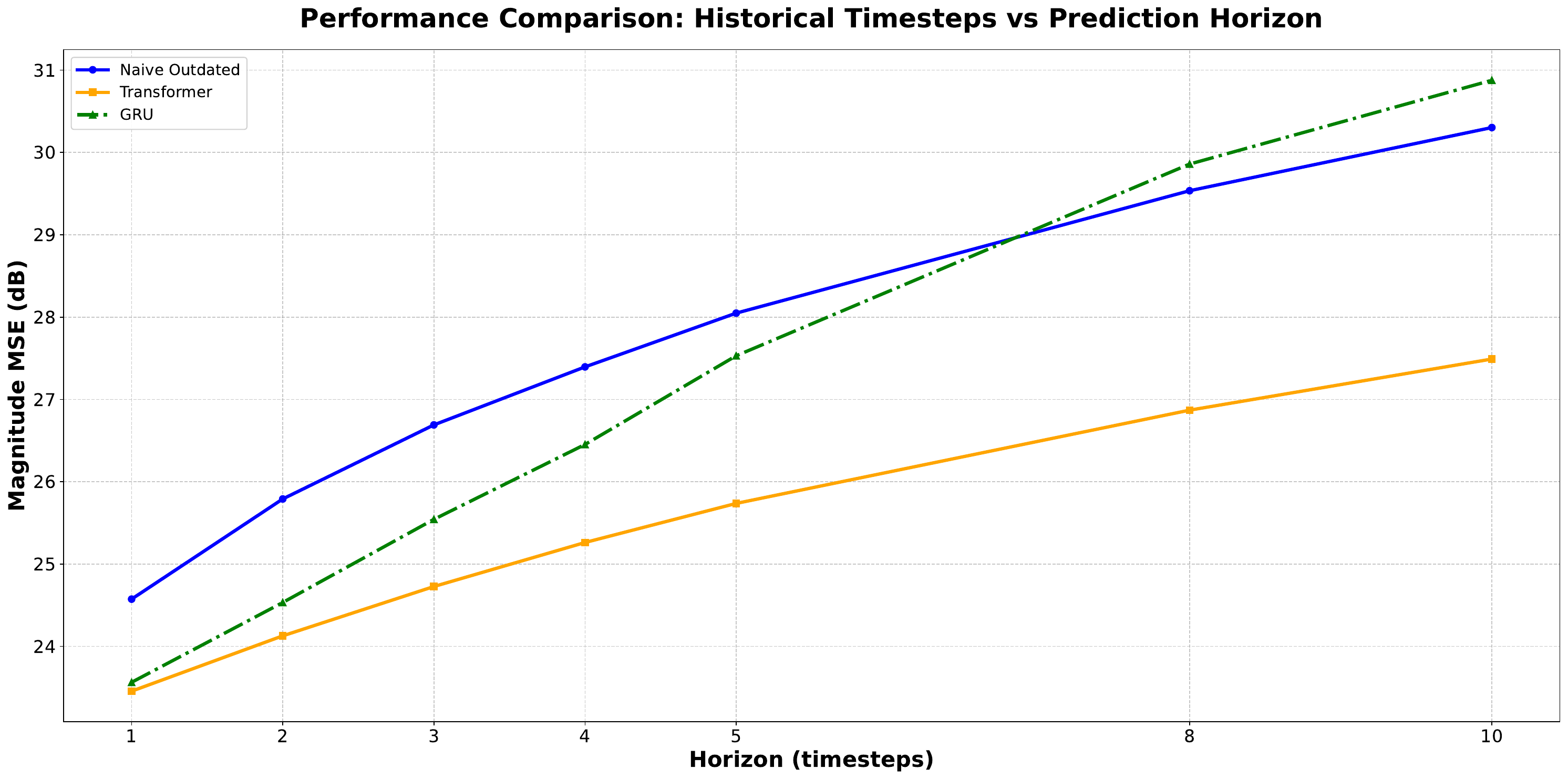}
    \caption{MSE (dB) vs. prediction horizon. Transformer shows superior long-term robustness compared to RNN baselines.}
    \label{fig:mse_comparison}
\end{figure}

Fig.~\ref{fig:mse_comparison} reports the MSE across horizons. At $k = 1$, the Transformer reaches \SI{-23.5}{dB}; by $k = 10$, the \gls{gru} degrades to \SI{-30.8}{dB}, even below the Naive baseline, while the Transformer remains at \SI{-27.5}{dB}. This \SI{3}{dB} gap indicates that exposure to authentic hardware artifacts improves robustness to stochastic phase noise.

\subsection{System-Level Retrospective Simulation}
To rigorously validate the practical gain, we developed a standalone retrospective simulator that strictly implements the link adaptation logic of the \gls{oai} PHY layer. Crucially, to avoid the pitfalls of idealized simulations, our evaluation is driven not by synthetic noise models, but by the \textit{empirical} noise variance and \gls{snr} values recorded synchronously with the channel estimates during the measurement campaign. This methodology explicitly preserves real-world hardware impairments, such as the \gls{sdr} non-linearities identified in Section~\ref{section3}, within the throughput calculation.

\subsubsection{MCS Selection Methodology}
The simulation proceeds in a slot-by-slot manner following the 3GPP standard~\cite{3gpp38214}. First, the model predicts the channel vector $\hat{\mathbf{h}}_{t+\tau}$ for the target horizon using the historical sequence. Subsequently, in the link adaptation phase, this predicted channel is combined with the most recently observed noise power to estimate the effective \gls{snr}, allowing the simulator to select the highest \gls{mcs} index that satisfies a \gls{bler} target of $\le \SI{10}{\percent}$ based on standard \gls{cqi} tables from the \gls{oai} codebase. Finally, the transmission outcome (ACK/NACK) is evaluated by applying the selected \gls{mcs} to the ground truth future channel $\mathbf{h}_{t+\tau}$ and the recorded noise level, where a transmission is deemed successful if the realized mutual information exceeds the threshold for the chosen \gls{mcs}.

\subsubsection{Statistical Analysis}
We evaluated the system performance over a high-mobility segment characterized by rapid channel fluctuations. In terms of reliability, the Transformer achieves a BLER of \SI{7.00}{\percent}, operating safely within the typical \SI{10}{\percent} target for initial transmission defined by 3GPP TS 38.214 to maximize spectral efficiency via \gls{harq}. Conversely, the \gls{gru} baseline reaches \SI{10.80}{\percent}, exceeding this target and indicating a higher risk of retransmission avalanches. Furthermore, regarding spectral efficiency, the Transformer leads with an average throughput of \SI{4.45}{Mbps}, surpassing both the reactive (\SI{4.39}{Mbps}) and \gls{gru} (\SI{4.29}{Mbps}) schemes. This performance maintains the narrowest gap to the optimal Oracle bound (\SI{4.77}{Mbps}), proving that the model’s prediction accuracy effectively translates into higher data rates even under real-world noise conditions.

\subsubsection{Visual Inspection and Discussion}

\begin{figure}[t!]
    \centering
    \subfloat[Time-domain Throughput Snapshot]{
        \includegraphics[width=0.95\linewidth]{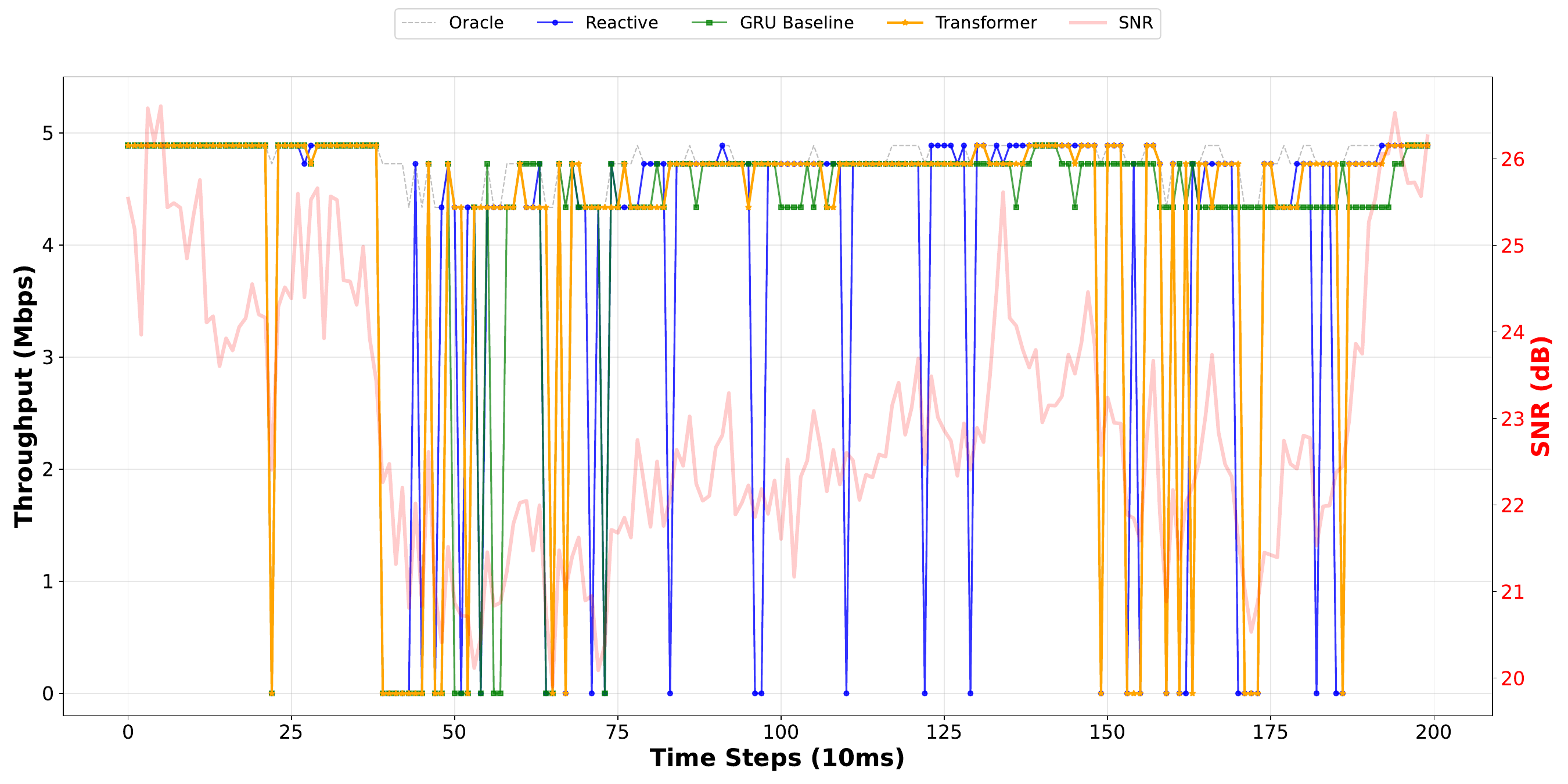}
        \label{fig:ts_snapshot}
    }
    
    \subfloat[Throughput CDF]{
        \includegraphics[width=0.9\linewidth]{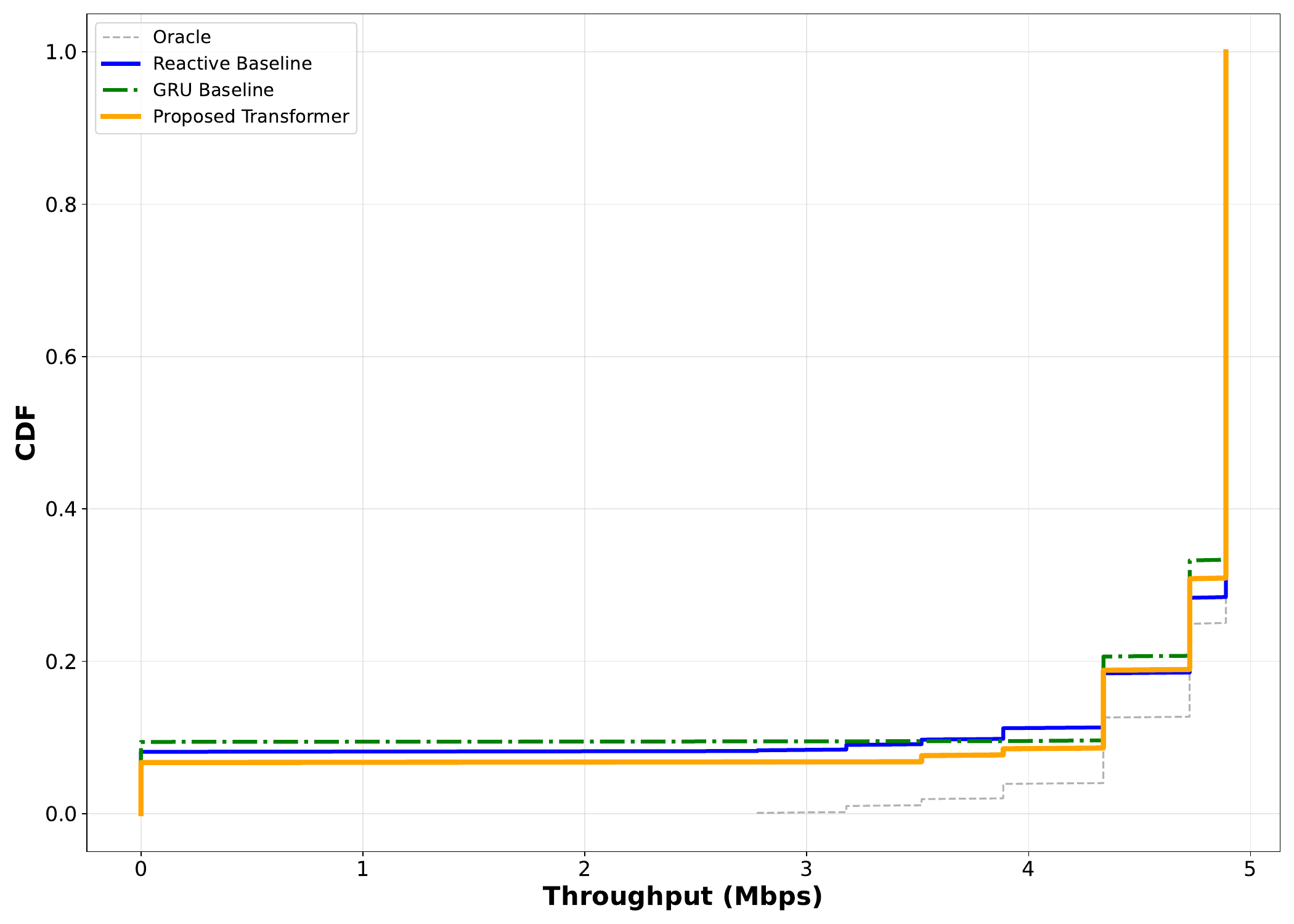}
        \label{fig:cdf_dist}
    }
    \caption{Link-level performance evaluation ($T_{\text{obs}}=10$, $T_{\text{pred}}=5$). (a) Time-domain throughput snapshot illustrating rate adaptation during deep fading events. (b) Throughput CDF comparison demonstrating the reliability distribution.}
    \label{fig:link_level_results}
\end{figure}

To understand the physical mechanism behind the Transformer’s superiority, Fig.~\ref{fig:link_level_results} presents a detailed visual comparison during a deep fading event.

As illustrated in the time-domain snapshot (Fig.~\ref{fig:ts_snapshot}), the channel conditions characterized by the red \gls{snr} line are highly volatile. Under these conditions, the naive baseline (blue) consistently reacts too late to sharp \gls{snr} drops, causing throughput to collapse to zero due to transmission failures. Similarly, the predictive \gls{gru} model struggles to distinguish between the underlying kinematic fading trend and the superimposed hardware phase noise; consequently, it often over-reacts to instantaneous noise spikes, resulting in highly unstable \gls{mcs} selection. In stark contrast, the Transformer (orange) proactively adjusts the data rate in anticipation of fading events.

The Transformer successfully tracks the true channel envelope, confirming its ability to filter the hardware jitter identified early in this section. The Cumulative Distribution Function (CDF) in Fig.~\ref{fig:cdf_dist} corroborates these findings. The Transformer curve (orange) strictly dominates the baselines, exhibiting the lowest packet error probability and the highest probability of achieving peak throughput.

\section{Conclusion}\label{section5}

To address the ``sim-to-real'' gap in AI-native \gls{6g} research, this paper presented a fully operational, open-source \gls{uav}-to-Ground measurement testbed utilizing \gls{oai} and commodity \glspl{sdr}. Our empirical spectral analysis revealed a clear velocity-bandwidth decoupling: in raw SDR measurements, the dominant observable spectral ridge does not directly follow the theoretical kinematic Doppler trajectory. Guided-versus-unguided ridge analysis further showed that Doppler information is present, but it is not the dominant component of the raw observable dynamics.

Furthermore, benchmarking on our dataset shows that models must be exposed to authentic hardware artifacts to filter non-kinematic jitter. The present study is limited to one open-source hardware chain, single-link UAV-to-ground measurements, and speeds up to \SI{15}{\meter\per\second}; the results therefore should not be over-generalized to higher-speed platforms, large-scale non-terrestrial networks, or multi-UAV systems. Future work will extend the benchmark to these regimes and compare against standardized simulated channel models and public datasets. By open-sourcing our testbed and dataset, we provide a realistic, hardware-inclusive foundation for evaluating robust AI models in future aerial networks.
\section*{Acknowledgements}
This work of the authors was supported by the German Federal Ministry of Research, Technology, and Space (BMFTR) through the 6GNeXt project (Grant no. \emph{16KISK177K}) and the Open6GHub+ project (Grant no. \emph{16KIS2402K}).

\bibliographystyle{IEEEtran}
\bibliography{bibliography.bib}

\end{document}